\documentclass[twocolumn, pra] {revtex4-1}    

\usepackage{graphicx}
\usepackage{dcolumn}
\usepackage{bm}

\renewcommand{\v}[1]{\vec{\mathbf{#1}}}

\begin{document}


\title{Engineering entanglement for metrology with rotating matter waves}

\author{L.M. Rico-Gutierrez}
\author{T.P. Spiller}
\author{J.A. Dunningham}
\affiliation{ School of Physics and Astronomy, University of Leeds, Leeds LS2 9JT, United Kingdom }


\begin{abstract}
Entangled states of rotating, trapped ultracold bosons form a very 
promising scenario for quantum metrology. In order to employ such states 
for metrology, it is vital to understand their detailed form and the 
enhanced accuracy with which they could measure phase, in this case 
generated through rotation. In this work we study the rotation of 
ultracold bosons in an asymmetric trapping potential beyond the lowest 
Landau level (LLL) approximation. We demonstrate that whilst the LLL can 
identify reasonably the critical frequency for a quantum phase 
transition and entangled state generation, it is vital to go beyond the 
LLL to identify the details of the state and quantify the quantum Fisher 
information (which bounds the accuracy of the phase measurement). We 
thus identify a new parameter regime for useful entangled state generation, 
amenable to experimental investigation.
\end{abstract}

\maketitle

%

\section{Introduction}
Quantum metrology is rapidly emerging as an exciting and feasible new technology based on quantum entanglement \cite{Holland1993a, Dunningham2002a, Gerry,Giovannetti2004,Dowling2008,Pezze2009a, Giovannetti2011, Jaewoo}. Great strides have been made in overcoming a range of practical issues that have hampered its implementation and proof of principle experiments have already been performed both with light \cite{Higgins2007, Kacprowicz2010a} and atoms \cite{Oberthaler} although real-world applications are yet to be realized \cite{Thomas-Peter2011a}. One class of systems that has been attracting particular attention is atomic Bose-Einstein condensates (BECs) trapped in rotating potentials \cite{fetter2009a}. These are appealing because they are within reach of current experiments \cite{Foot2006a, Boshier2009a, Ryu2007a}, provide a conceptually simple way of generating many-body entanglement, and have the prospect of leading to the development of ultra-precise gyroscopes \cite{Cooper2010a}.

A particularly promising system for achieving these goals is a BEC in a quasi two-dimensional trap with a weak anisotropic stirring potential. As the rotation frequency is slowly ramped up from zero, the ground state passes a critical frequency $\Omega_\mathrm{c}$ signalling the emergence of the first vortex. At this point, the system undergoes turbulent symmetry breaking that heralds a quantum phase transition and the many-body state is a strongly correlated entangled state involving two macroscopically occupied modes \cite{dagnino, Gemelke}. Such systems may be amenable to experimentally generating entangled states for use in quantum metrology schemes. 

In this paper, we present a detailed calculation to assess the usefulness of the entangled states that can be created. We begin with a LLL calculation. This allows for a simplified and more computationally tractable description of the system but requires that the particle density is low, the interaction strength is small, or the rotation frequency is very close to the harmonic trap frequency.  We find that this approximation accurately predicts the critical frequency $\Omega_\mathrm{c}$ to within a few percent. This could lead us to believe that it provides a good description of the system. However a deeper investigation that goes beyond LLL reveals that this is not true in general. We find, for example, that the `true' many-body state  at $\Omega_\mathrm{c}$ and the quantum Fisher information, $F_Q$, which quantifies how well a state performs in a quantum metrology scheme, are significantly different from the LLL predictions.

So, while the LLL approximation may be sufficient for some purposes such as studying symmetry breaking \cite{dagnino}, a more detailed calculation that includes a larger basis is vital if we are interested in the precise details of the entangled state or if we want to use it in applications. This issues a warning about the use of the LLL approximation, particularly as a guide for designing experiments. A study of the anisotropic trap with our more detailed calculation reveals a rich system with interesting opportunities for engineering different entangled states. In particular, we are able to identify a new parameter regime for generating entanglement with a form very similar to a bat state \cite{Holland1993a, Dunningham2005a}. Bat states are known to be well-suited to metrology because they combine high measurement precision with robustness to loss. This gives them a distinct advantage over N00N states, which are the archetypal states for metrology. N00N states have the form $(|N\rangle|0\rangle + |0\rangle|N\rangle)/\sqrt2$, where the two kets in each term represent the number of particles in two different states, and are very fragile because as soon as there is any information about which state any one particle is in, the superposition is destroyed.

\section{Model}
The physical system we consider consists of a mesoscopic sample of $N$ bosonic atoms of mass $M$ in an axially symmetric harmonic potential, with frequency $\omega_\perp$ in the $xy$ plane and $\omega_z$ in the $z$ axis, interacting through hard-core-type elastic collisions. We take $\hbar \omega_z$ to be very large compared to the interaction energy so that the dynamics along the $z$ axis are frozen, i.e. all particles occupy the lowest axial energy level, thus rendering the gas effectively two-dimensional at sufficiently low temperatures. The trapped gas is rotated at angular frequency $\Omega$ around the $z$ axis with the aid of an external potential which in the rotating frame appears as an anisotropic quadratic potential in the $xy$ plane. The Hamiltonian in the rotating frame is given by 
\begin{widetext}
\begin{equation}
H=\sum_{i=1}^N-\frac{\hbar^2}{2M}\nabla_i^2+\frac{1}{2}M\omega_\perp^2(x_i^2+y_i^2)+2A M\omega_\perp^2(x_i^2-y_i^2)-\Omega L_{zi}+\frac{\hbar^2 g}{M}\sum_{j<i}^N \delta(\v{r}_j-\v{r}_i),
\label{theH}
\end{equation}
\end{widetext}
where $A(\ll 1)$ quantifies the degree of anisotropy, which is treated perturbatively in the calculations. Here, $L_{zi}$ is the angular momentum component in the $z$ direction of the $i$-th atom and the  $-\Omega L_z$ term transforms the Hamiltonian to the rotating frame, where $L_z$ is the total angular momentum of the condensate. Finally, $g$ is the dimensionless interaction coupling constant which quantifies the strength of two-particle interactions and is related to the 3D scattering length $a$ by \mbox{$g=\sqrt{8\pi} a / \lambda_z$}, where \mbox{$\lambda_z=\sqrt{\hbar/M\omega_z}$}.

In the absence of anisotropy ($A=0$), the Hamiltonian is exactly diagonalizable in blocks of definite total angular momentum by choosing a Fock basis of the form
\begin{equation}
|N_0, N_1, \ldots \rangle = \prod_\mathbf{k} \frac{\hat{a_\mathbf{k}}^{N_\mathbf{k}}}{\sqrt{N_\mathbf{k}!}} |0\rangle,
\label{theBasis}
\end{equation}
where $\hat{a_\mathbf{k}}$ creates a boson in the single-particle state $\mathbf{k}$ and $N_\mathbf{k}$ is the occupation number of level $\mathbf{k}$. Here, the index $\mathbf{k}$ represents a pair of quantum numbers $(n,m)$, the principal quantum number and the projection of the angular momentum respectively \cite{feder}. For this basis, in the strictly noninteracting case, the Hamiltonian is diagonal with eigenvalues \mbox{$E/\hbar\omega_\perp=2\sum_i^N n_i+\sum_i^N |m_i| -\Omega L + N$}, where $L$ is the total angular momentum of the system in units of $\hbar$ and $\Omega$ is the rotation frequency in units of $\omega_\perp$. Therefore, in the case of independent bosons, when $\Omega\approx\omega_\perp$ the spectrum consists of highly degenerate levels called Landau levels that are separated by an energy gap of $2\hbar\omega_\perp$. 

The LLL approximation corresponds to considering only the states with $n_i=0$ and $m_i\ge 0$. This greatly reduces the Hilbert space dimension, allowing the simulation of larger numbers of atoms, as well as being very convenient for computational implementations. The LLL approximation is valid when the chemical potential is much smaller compared to the level spacing, i.e. $\mu\ll 2\hbar\omega_\perp$. In the limit of fast rotations, $\Omega\rightarrow\omega_\perp$, the chemical potential is just the mean-field interaction energy per particle, i.e.  $\widetilde{g} \, \overline{n}$, where $\overline{n}$ is the mean density of the condensate and $\widetilde{g}$ is the interaction strength related to the dimensionless coupling constant by \mbox{$\widetilde{g}=\hbar^2 g/M$}. Therefore, in this limit, the standard validity criterion is that $\widetilde{g} \, \overline{n}\ll 2\hbar\omega_\perp$.

 In the present study, we will investigate the effect of including more Landau levels in the calculations, when doing so, we define the $n_\mathrm{LL}$-th Landau level approximation as that one obtained by including only basis states where \mbox{$\sum_{i=1}^{N}\left[n_i+(|m_i|-m_i)/2\right]=n_\mathrm{LL}-1$}, as considered in \cite{feder}.

When $A\neq 0$ the anisotropic term connects subspaces of total angular momentum separated by two units of $\hbar$. If the anisotropy is small, a finite maximum angular momentum $L_\mathrm{max}$ can be chosen to ensure convergence for the energies and eigenstates of the the Hamiltonian restricted to the subspace of states with $L\le L_\mathrm{max}$ as considered in \cite{dagnino}. Throughout the rest of this paper, we use $\omega_\perp$ as units of frequency and $\sqrt{\hbar/(M\omega_\perp)}$ as units of length.

\begin{figure}
\includegraphics[width=0.43\textwidth]{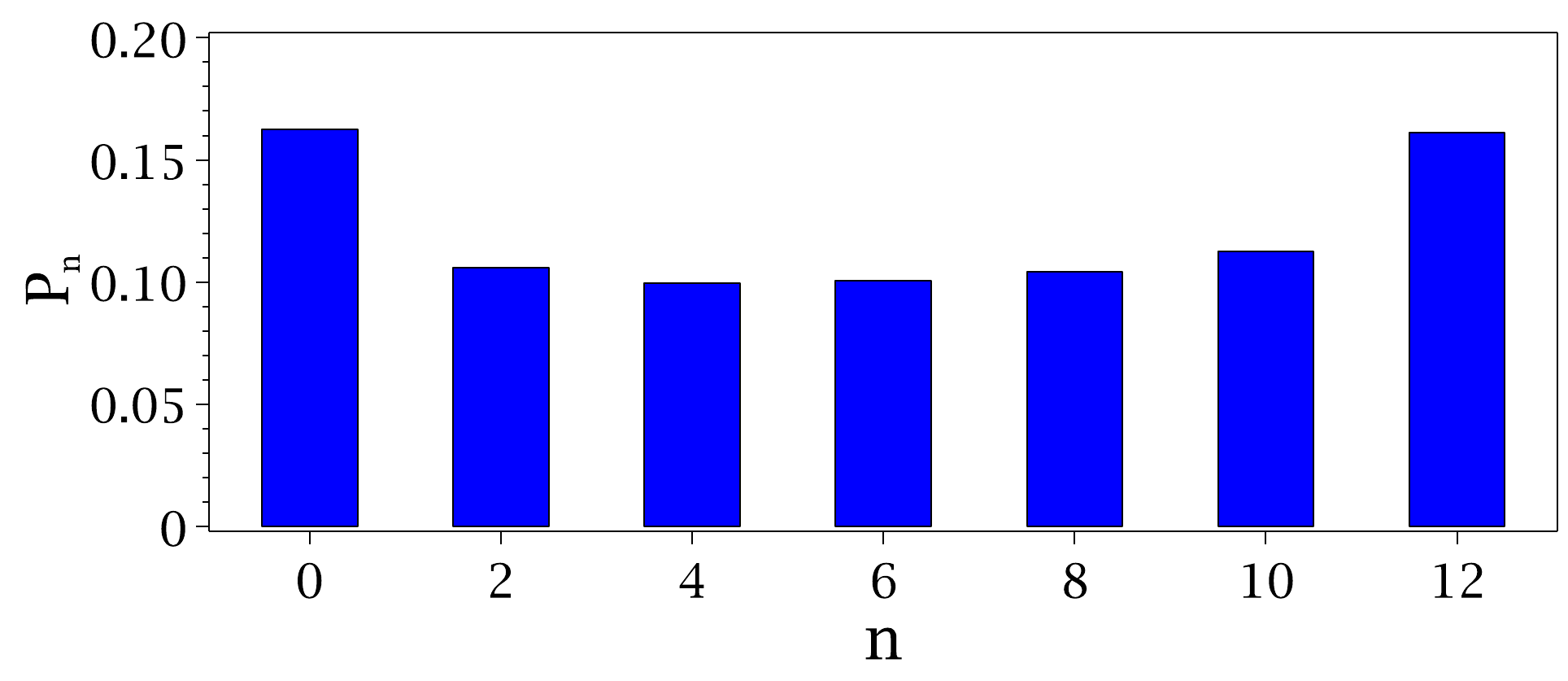}
\includegraphics[width=0.43\textwidth]{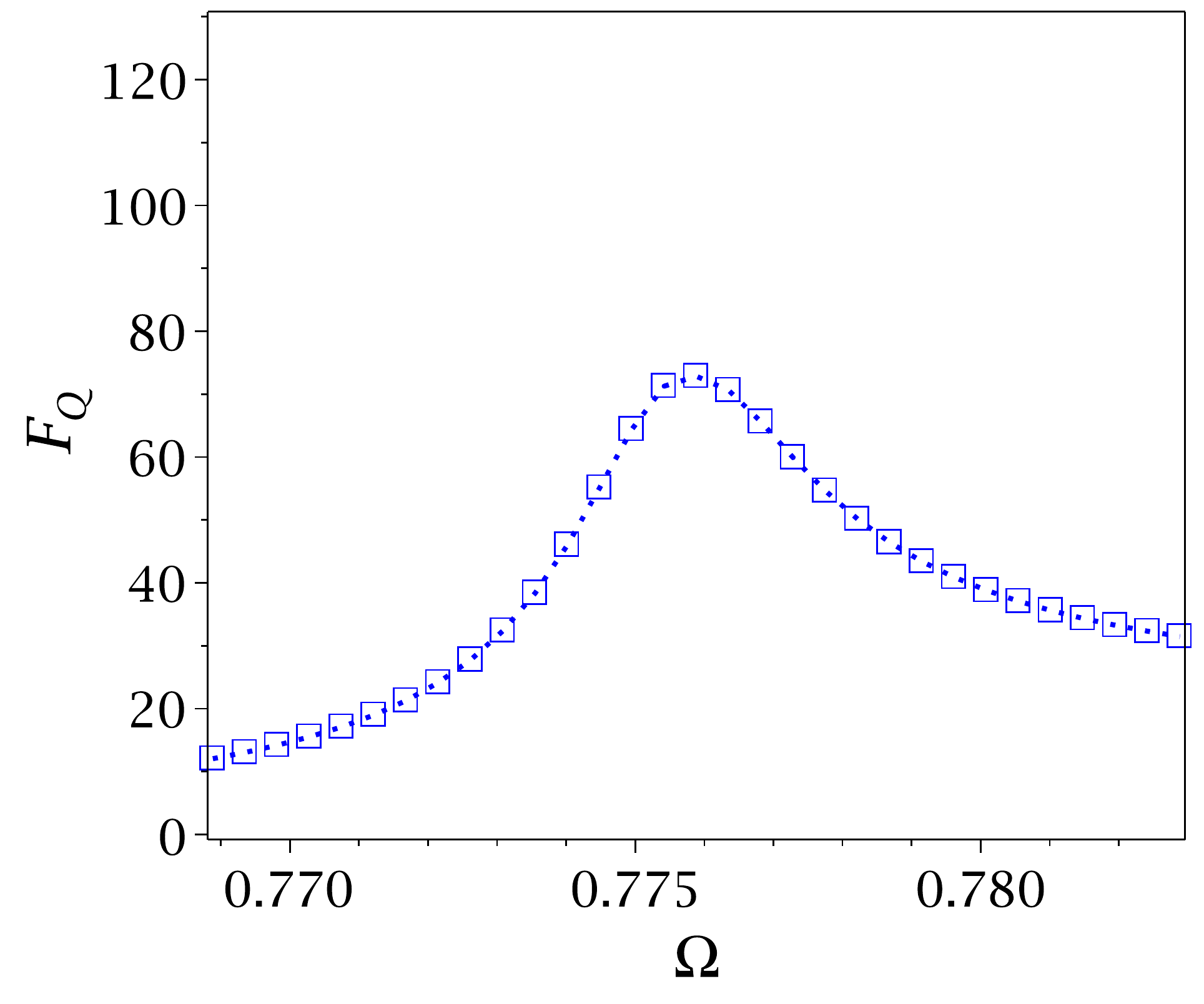}
\caption{\label{fig:LLL} Analysis of the ground state at the critical point using the lowest Landau level approximation for $N=12$, $g=0.5$ and $A=0.03$. (Top) The coefficients \mbox{$P_n=|\langle N-2n, 2n | \Psi_\mathrm{GS} \rangle|^2$} for the two-mode approximation to the many-body ground state at \mbox{$\Omega_c\approx 0.776$}. The fidelity of the two-mode approximation is found to be \mbox{$|\langle \Psi_\mathrm{TM} | \Psi_\mathrm{GS} \rangle|^2=0.83$}. (Bottom) The quantum Fisher information as a function of rotation rate $\Omega$. }
\label{fig1}
\end{figure}

\section{Results} 
The case considered in \cite{dagnino} is that of a relatively slowly rotating gas at the threshold of the nucleation of the first vortex, where the strength of the interaction and the anisotropic perturbation are considered sufficiently small compared to the separation of the Landau levels in order to assume the LLL regime. In particular, the anisotropy strength is $A=0.03 \ (A\ll 1)$ and the interaction one is assumed to be $g \ N=6$ for all the cases considered therein with different number of particles. This last expression is consistent with the standard mean-field criterion for the validity of the LLL approximation, $\mu \ll 2\hbar\omega_\perp$. When $N g=6$ the typical rotational frequency for the nucleation of the first vortex  is $\Omega>0.77$ and it can be shown that in unitless form \cite{fetter2010}, the standard criterion for the LLL validity is equivalent to $N g\ll 2\pi/(1-\Omega)$, hence the $N g=6$ scaling is consistent with the LLL approximation. More importantly, it is also consistent with the power law scaling $g_\mathrm{max}=(6.92\pm 0.08) N^{-1.046\pm 0.005}$ for small number of particles $(N<20)$ describing the crossover from weakly to strongly interacting regimes for the first vortex nucleation as found by Feder \cite{feder}.
 
As the rotation frequency is adiabatically swept, the system changes from being at rest to containing one vortex. In doing so, the ground state of the system passes a critical frequency $\Omega_c$ where the system undergoes turbulent symmetry breaking indicating a quantum phase transition.


Moreover, for even numbers of particles, the approximation also predicts that at the critical frequency $\Omega_c$, the ground state is a strongly correlated entangled state well-described by a two-mode approximation
\begin{equation}
|\Psi_\mathrm{TM}\rangle = \sum_{n=0}^{N/2} C_n |N-2n\rangle|2n\rangle,
\label{twomodeGS}
\end{equation}  
where \mbox{$|N-2n\rangle|2n\rangle$} is the state with \mbox{($N-2n$)} (respectively \mbox{($2n$)}) bosons in the the most (second most) populated single-particle state $\psi_1$ ($\psi_2$). These two states are eigenstates of the single-particle density matrix (SPDM) \mbox{$\rho^{(1)}_{l k}=\langle \Psi_\mathrm{GS}| \hat{a}_\mathbf{k}^\dagger \hat{a}_\mathbf{l} |\Psi_\mathrm{GS}\rangle$} and their populations are equal at the critical frequency, which together account for most of the population of the SPDM. Below the critical frequency, the most populated single-particle state $\psi_1$ is approximately a coherent superposition of two off-centred vortices with even parity corresponding to the states with $m=0$ units of angular momentum and $m=1$ units. The second most populated state is a well-centred single-vortex state with odd parity corresponding to the state with $m=1$. Above the critical frequency, $\psi_1$ and $\psi_2$ abruptly swap places.
 
The relevant results using the LLL approximation (as in \cite{dagnino}) for the current discussion are shown in Fig.\ref{fig:LLL} for $N=12$ particles. The top plot shows the coefficients $P_n=|C_n|^2$ for the two-mode approximation of the many-body ground state in Eq.(\ref{twomodeGS}). This is promising for quantum metrology since the state has a ``bat-like'' structure which is known to be robust to particle loss in interferometric schemes \cite{Cooper2010a}. The bottom plot is the quantum Fisher information as a function of rotation rate $\Omega$. This is a measure of how well the quantum state can perform in metrology schemes. Suppose that the state given by Eq.(\ref{twomodeGS}) picks up an undetermined linear phase as a result of a unitary operation $U(\phi)$ being applied to the first mode, i.e. 
\begin{equation}
U(\phi)|\Psi_\mathrm{TM}\rangle = \sum_{n=0}^{N/2} C_n e^{i\phi(N-2n)}|N-2n\rangle|2n\rangle.
\label{GSphase}
\end{equation}     
The best precision, $\Delta\phi$, with which $\phi$ can be measured using this state in a single shot experiment, independent of the details of the measurement scheme, is given by the quantum Cram\'er-Rao bound \cite{Braunstein1996},
\begin{equation}
\Delta\phi \ge \frac{1}{\sqrt{F_Q}},
\label{CramerRao}
\end{equation} 
where $F_Q$ is the quantum Fisher information. For a pure state $|\Psi(\phi)\rangle$ that depends on a single parameter $\phi$, $F_Q$ is given by
\begin{equation}
F_Q=4\left[\langle \Psi'(\phi) | \Psi'(\phi)\rangle - |\langle \Psi'(\phi)|\Psi(\phi)\rangle|^2\right],
\label{QFIpure}
\end{equation}             
where $|\Psi'(\phi)\rangle=\partial|\Psi(\phi)\rangle/\partial\phi$. The results for the Fisher information in Fig.\ref{fig:LLL} show a relatively broad curve in the rotation frequency, which is a desirable property because it implies that experimentalists would have a sizeable margin of error when trying to hit the critical frequency to prepare the initial entangled state.

\begin{figure}
\includegraphics[width=0.43\textwidth]{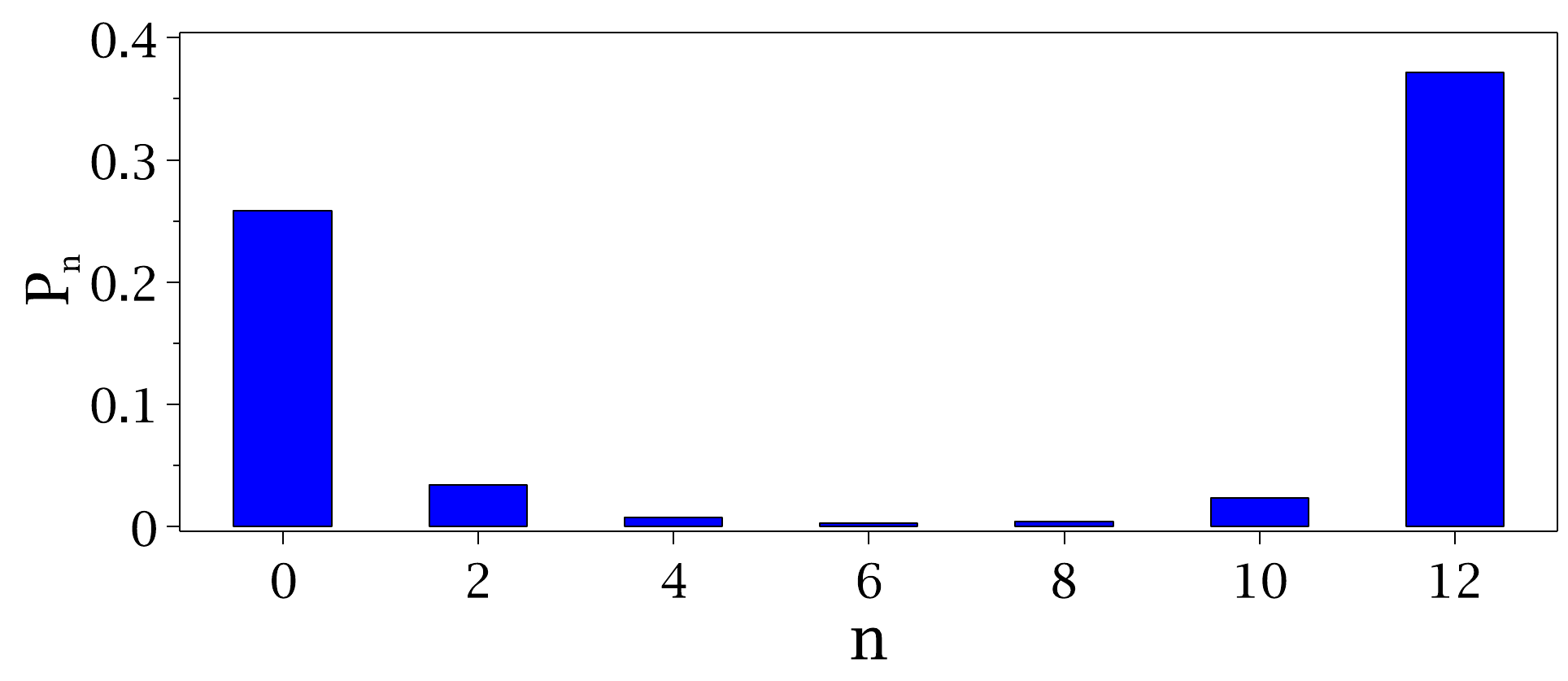}
\includegraphics[width=0.43\textwidth]{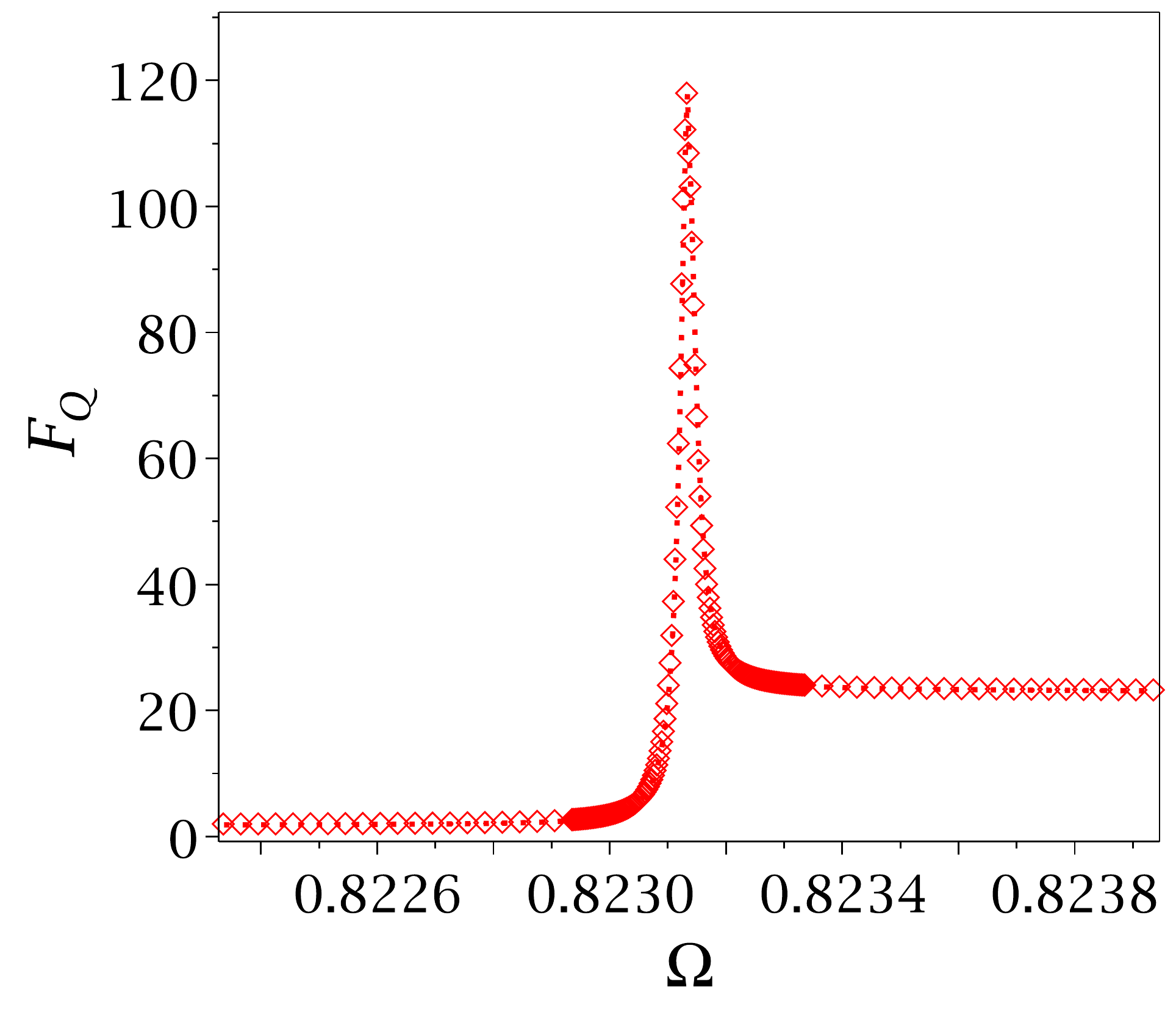}
\caption{\label{fig:twoLL} Analysis of the ground state at the critical point using a two Landau level approximation for $N=12, g=0.5$ and $A=0.03$. (Top) The coefficients \mbox{$P_n=|\langle N-2n, 2n | \Psi_\mathrm{GS} \rangle|^2$} for the two-mode approximation to the many-body ground state at \mbox{$\Omega_c\approx 0.823$}. The fidelity of the two-mode approximation is found to be \mbox{$|\langle \Psi_\mathrm{TM} | \Psi_\mathrm{GS} \rangle|^2=0.70$}. (Bottom) The quantum Fisher information as a function of rotation rate $\Omega$. Notice that the scale is roughly 10 times smaller than that of Fig.~\ref{fig:LLL}.}
\end{figure}

Surprisingly, the form of the ground state and quantum Fisher information around criticality change dramatically when another Landau level is taken into account in the calculations. The results for the same set of parameters but using two Landau levels are shown in Fig.\ref{fig:twoLL}. Although the critical frequency has been shifted by roughly 5\%, the ground state at the critical frequency has now become more like a N00N state, $(|N\rangle|0\rangle + |0\rangle|N\rangle)/\sqrt2$, which can attain Heisenberg-limited precision that scales as $\Delta\phi\sim 1/N$ but, unlike a ``bat-like'' state, is extremely fragile to particle loss. The fidelity of the ground state in the two-mode approximation with the full ground state is also reduced due to the increasing participation of a third single-particle mode which was not present in the LLL results. Furthermore, the quantum Fisher information shows an equally striking difference compared to the LLL approach. The Fisher information curve has become very sharp: its width has been reduced by about two orders of magnitude compared with Fig.~\ref{fig1}. This sharp resonance means that it is likely to be experimentally very challenging to produce a N00N state because a high control of the rotation frequency is required in order to hit the critical frequency exactly. Here, it is important to mention that the quantum Fisher information has been calculated using the full ground state as obtained from the exact diagonalization procedure and after a single-particle change of basis to the natural orbitals one (for which the SPDM is diagonal), rather than using the two-mode approximation for the ground state. 


As it stands, these results do not look particularly promising for metrology. Fortunately, we can get all the benefits that the results in \cite{dagnino} suggest by looking more widely in parameter space. This is done in the next section where we assess the validity of the LLL approximation in the current context.

\section{Validity of the LLL}
We have found in the previous section that, although the scaling $N g_\mathrm{max} = 6$ provides a value for the maximum interaction strength $g_\mathrm{max}$ that is consistent with the use of the LLL approximation to accurately predict the critical rotation frequency $\Omega_c$ for a small number of atoms $(N<20)$, it is not a good guideline to correctly predict the quantum Fisher information or the exact form of the multi-body ground state at the critical frequency. This scaling relation is the result of thorough numerical evidence for axisymmetric traps, where the criteria used to establish the validity of the LLL approximation is that the maximum interaction strength $g_\mathrm{max}$ predicts the rotation frequency $\Omega_1$ for the nucleation of the first vortex within a relative error of 10\% between the results of the LLL aproximation and those obtained with two or three Landau levels \cite{feder}. At this nucleation frequency $\Omega_1$, the ground state and first excited state become degenerate and their energies show a crossing point. Therefore, in order to bring the condensate from the non-rotating ground state to the ground state containing one vortex, it is neccesary to break the rotational symmetry, thus creating an avoided crossing that allows the passage of the system to the rotating regime with total angular momentum $L=N$ and one vortex. The rotational symmetry can be broken using a weak ($A\ll 1$) anisotropic stirring potential that helps to both impart angular momentum to the system and create the avoiding crossing. This approach has been achieved experimentally for a few number of atoms in optical lattices, where the atoms in each lattice are set in rotations using a local anisotropic stirring potential \cite{Gemelke}.  Naturally, it is reasonable to expect that the addition of this very small anisotropic perturbation ($A\ll 1$) still allows for the description of the system in terms of the axisymmetric solutions. However, surprisingly enough, no matter how small $A$ is (we checked this for values as small as of $A=1\times 10^{-3}$, which is smaller than the typical static anisotropy defect in real traps), the striking differences in the quantum Fisher information and the exact form of the ground state found in the previous section still hold. Consequently, it is important to investigate the validity of the LLL approximation in the current context of the inclusion of small anisotropic perturbation. Some intuition as to the reason why the LLL approximation has failed to capture the physics predicted by the addition of more Landau levels can be obtained by taking a closer look at the energy spectrum of the many-body Hamiltonian $\hat{H}$. In the absence of the anisotropic term ($A=0$), the interacting many-body Hamiltonian is exactly diagonalizable in blocks of definite angular momentum since the interaction term commutes with the total angular momentum ($\hat{L}_z$ in the present case of a 2D gas) and therefore does not mix subspaces of different angular momentum, consequently, the eigenvectors of the interacting Hamiltonian are linear combinations of angular momentum eigenstates with the same eigenvalue of $\hat{L}_z$. In the LLL approximation, all the lowest eigenstates of $\hat{H}$ with $L=0,2,\ldots N$ become degenerate at the first energy crossing regardless of the interaction strength $g$ and the ground state and first excited state consist of the states with $L=0$ and $L=2$ respectively for rotation frequencies below and near the frequency of the crossing and the states $L=N$ and $L=N-2$ for frequencies above and near the frequency of the crossing. The effect of adding a small anisotropic term to $\hat{H}$ is that of lifting the degeneracy and nucleating an entagled ``bat-like'' state at the critical frequency $\Omega_c$ which is mainly a superposition of the ground state $L=N$ and excited state $L=N-2$ of the unperturbed Hamiltonian. These two states are directly connected by the perturbation and they are energetically close together, thus making the superposition favorable. In contrast, when another Landau level is included, the isotropic eigenstates with $L=0,2,\ldots N$ no longer cross at the same frequency, however, they remain quasi-degenerated for small values of \mbox{$g\sim (6/N)\times 0.4$} and thus the addition of the anisotropic term has the same effect as in the LLL case, i.e. the nucleation of a ``bat-like'' state. In fact, the entangled ground state at $\Omega_c$ is again mainly a superposition of the states $L=N$ and $L=N-2$. On the other hand, when the interaction strength becomes of the order of $g\sim (6/N)$, the eigenstate $L=0$ in the isotropic case has a greater energy separation from the other $L=2\ldots N$ states which remain fairly quasi-degenerate at a certain frequency which is no longer the first energy crossing for the ground state, instead, the first crossing occurs above this frequency and consists only of the states with $L=0$ and $L=N$. In this case, the addition of the anisotropic term results in the nucleation of a ``cat-like'' state which is now mainly a superposition of the states $L=0$ and $L=N$. Although these states are not directly connected by $A$, their proximity in energy and the fact that the states to which they are directly connected through the anisotropy are farther away in energy, favors such superposition.

\begin{figure}
\includegraphics[width=0.43\textwidth]{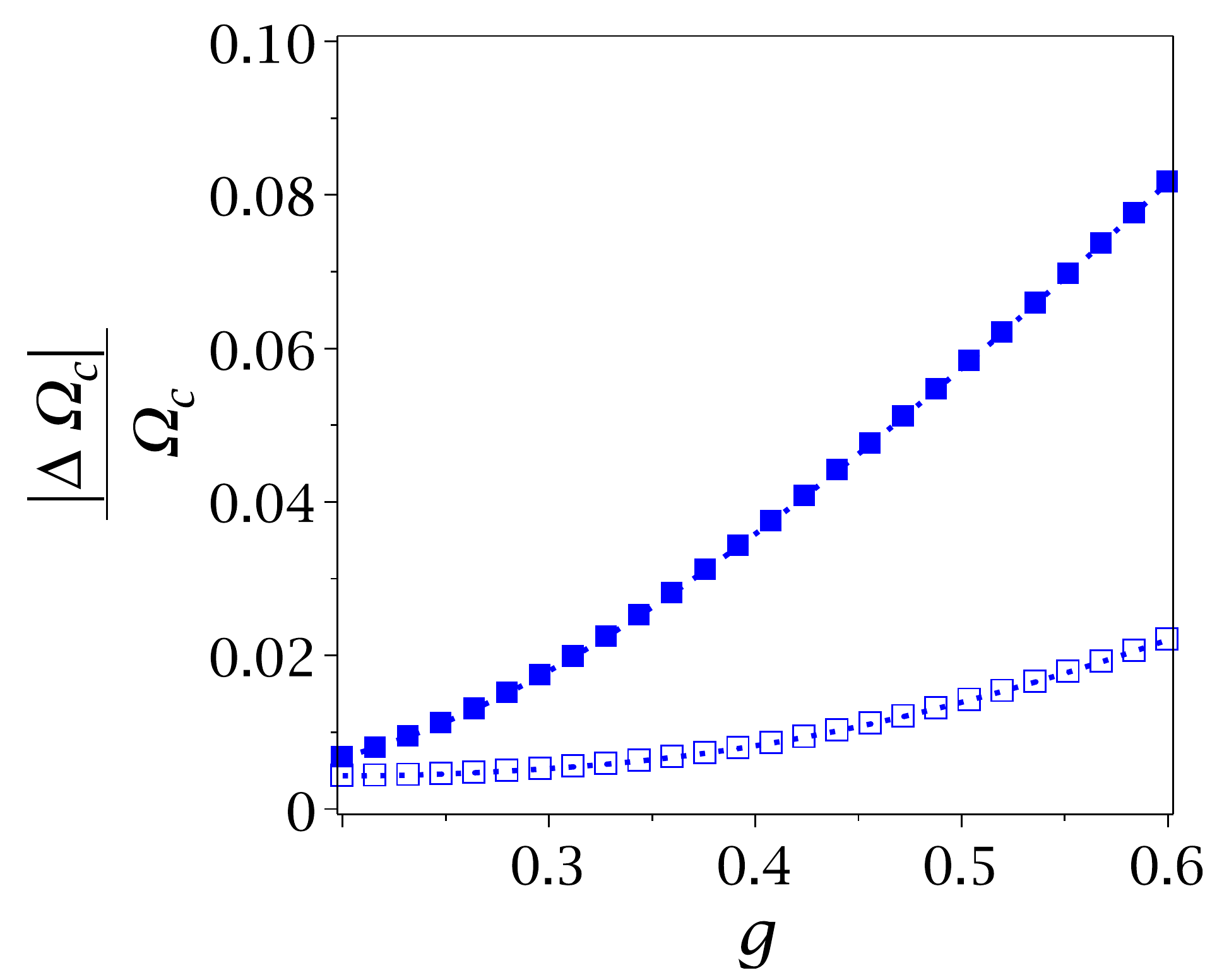}
\caption{\label{fig:FracCF} The fractional change in the position of the critical rotation frequency $\Omega_c$ as a function of the interaction strength, calculated using two different approximations differing by one Landau level. In each case, the fractional change is calculated at the critical frequency obtained using the lower number of Landau levels. One level approximation (LLL) compared against the two level approximation (solid box) where \mbox{$\Delta \Omega_c / \Omega_c=|\Omega_c(g)_\mathrm{2LL}-\Omega_c(g)_\mathrm{LLL}|/\Omega_c(g)_\mathrm{LLL}$}. Two level approximation compared against the three level approximation (empty box) where \mbox{$\Delta \Omega_c / \Omega_c=|\Omega_c(g)_\mathrm{3LL}-\Omega_c(g)_\mathrm{2LL}|/\Omega_c(g)_\mathrm{2LL}$}. We only show results for $g>0.2$ since below this value, the participation of a rapidly increasing third single-particle mode deteriorates the fidelity of the ground state in the two-mode approximation and shifts the critical frequency above the trapping frequency which is physically unattainable. } 
\end{figure}

\begin{figure}
\includegraphics[width=0.43\textwidth]{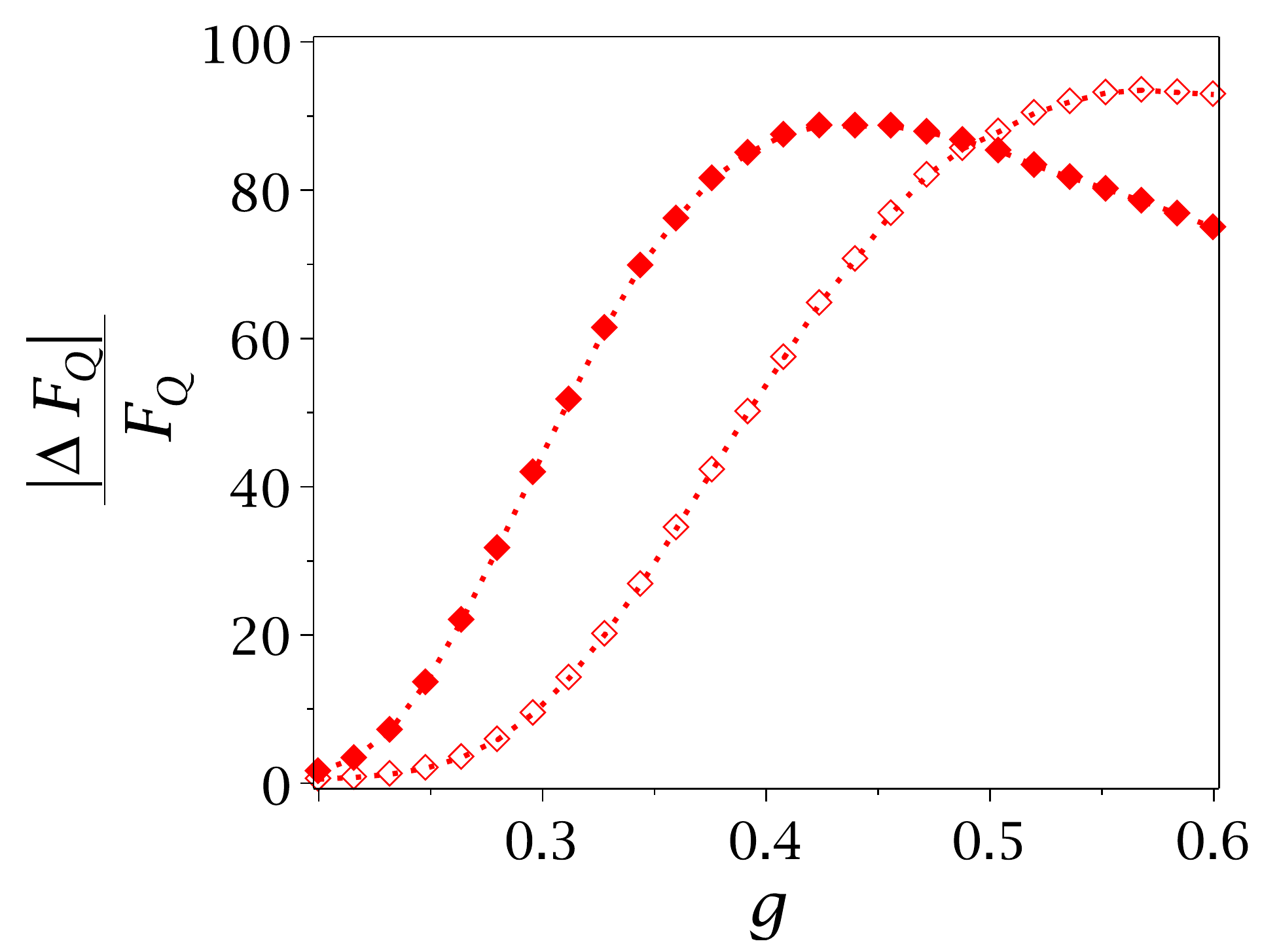}
\caption{\label{fig:FracFq} The fractional change in the quantum Fisher information $F_Q$ as a function of the interaction strength, calculated using two different approximations differing by one Landau level. In each case, the fractional change is calculated at the critical frequency obtained using the lower number of Landau levels. One level approximation (LLL) compared against the two level approximation (solid diamond) where \mbox{$\Delta F_Q / F_Q=|F_Q(g)_\mathrm{2LL}-F_Q(g)_\mathrm{LLL}|/F_Q(g)_\mathrm{2LL}$}. Two level approximation compared against the three level approximation (empty box) where \mbox{$\Delta F_Q / F_Q=|F_Q(g)_\mathrm{3LL}-F_Q(g)_\mathrm{2LL}|/F_Q(g)_\mathrm{3LL}$}. } 
\end{figure}

\begin{figure}
\includegraphics[width=0.43\textwidth]{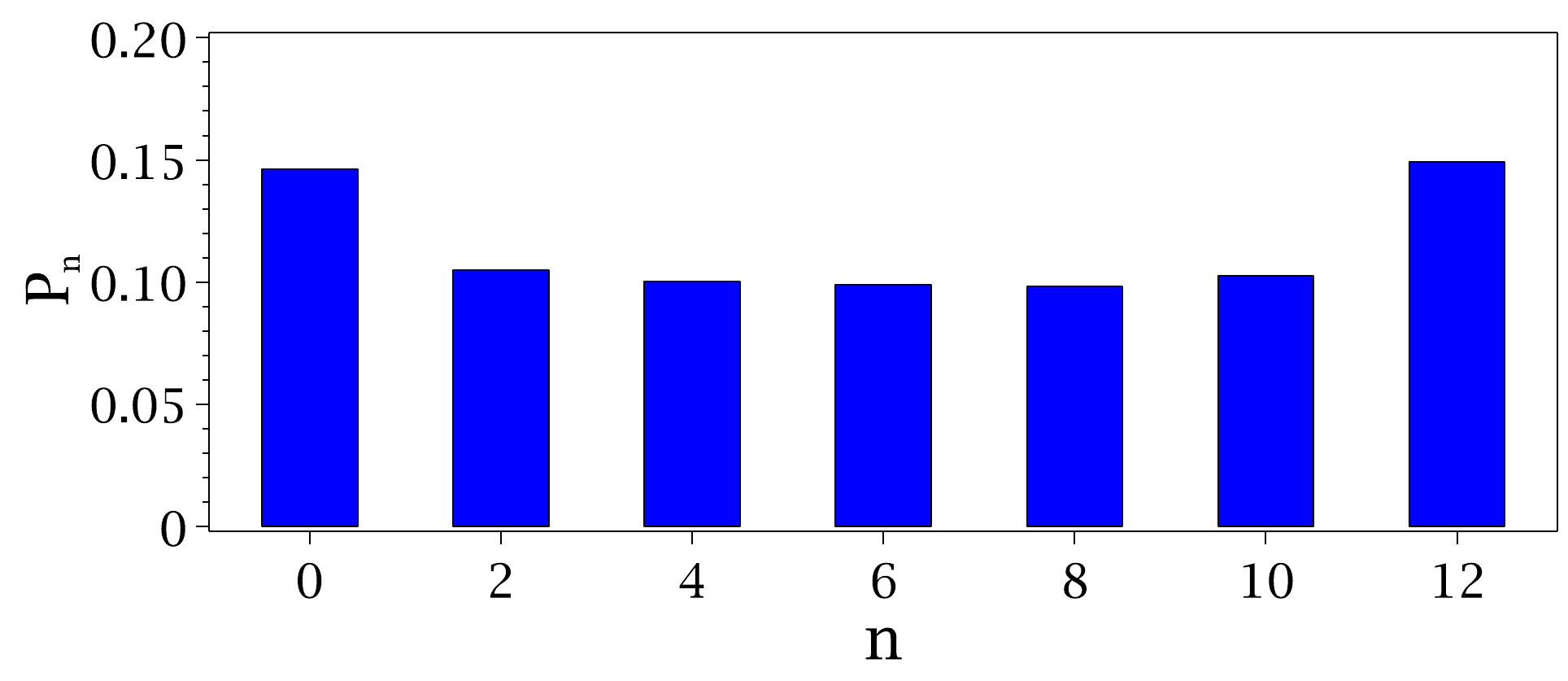}
\includegraphics[width=0.43\textwidth]{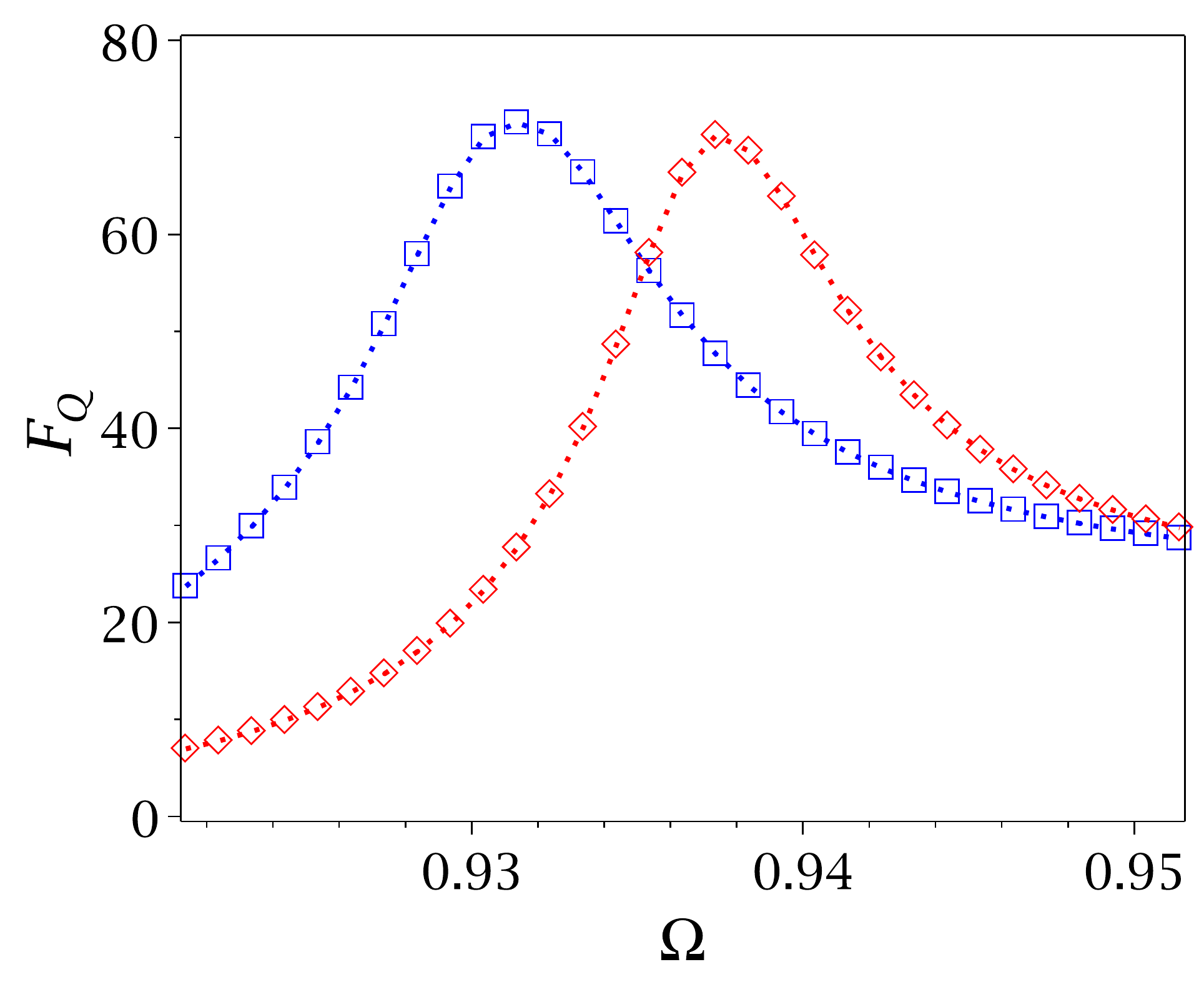}
\caption{\label{fig:sweetspot} Analysis of the ground state at the critical point using a two Landau level approximation for an interaction strength of $g=0.2$ and $N=12$, and $A=0.03$, which has been tuned for better suitability of the ground state for quantum metrology. (Top) The coefficients \mbox{$P_n=|\langle N-2n, 2n | \Psi_\mathrm{GS} \rangle|^2$} for the exact ground state at \mbox{$\Omega_c\approx 0.938$} calculated with the two level approximation. The fidelity of the two-mode approximation is found to be \mbox{$|\langle \Psi_\mathrm{TM} | \Psi_\mathrm{GS} \rangle|^2=0.80$}. (Bottom) The quantum Fisher information as a function of rotation rate $\Omega$ for the LLL approximation (blue empty box) and two Landau levels (red empty box).}
\end{figure}

\begin{figure}
\includegraphics[width=0.43\textwidth]{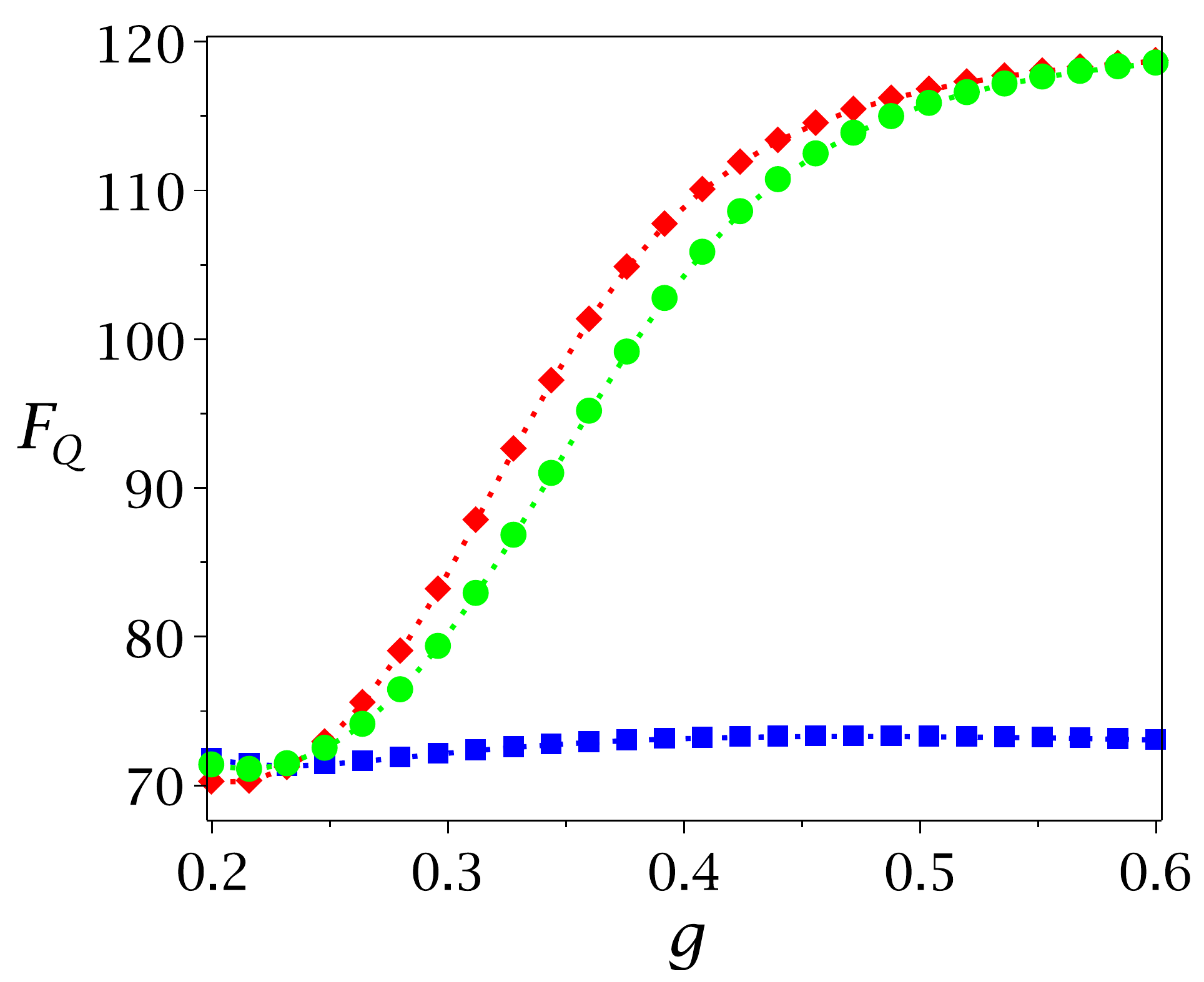}
\caption{\label{fig:Fq} The quantum Fisher information of the ground state for various basis truncations as a function of the interaction strength at the critical frequency $\Omega_c$ calculated with the respective truncated basis.Three different basis truncation are considered: one Landau level (blue solid box), two Landau levels (red solid diamond), and three Landau levels (green solid circle). }  
\end{figure}

In order to further quantify the validity of the LLL approximation, we have considered the fractional change in two different quantities as given by the calculations performed with the LLL approximation and those obtained with more Landau levels included for $N=4,6,8,10$ and $12$ particles. The results are qualitatively similar  for all the studied number of particles and we show the ones for $N=12$ for definiteness. the ones for $N=12$ for definiteness. First, we consider the predicted critical frequency $\Omega_c$. Our results shown in Fig.\ref{fig:FracCF} are consistent with the scaling $N g_\mathrm{max}=6$ for axisymmetric traps. This critical rotational frequency can be predicted within a relative error of less than 10\% when using the LLL approximation and the two level approximation for a broad range of values of the interaction strength. Moreover, an extremely accurate prediction of the critical frequency is given by a two level approximation compared to the one obtained with three Landau levels. This picture is completely similar to the one that happens in axisymmetric traps when the fractional change in $\Omega_1$ is considered. A totally different picture is observed for the fractional change of the quantum Fisher information in Fig.\ref{fig:FracFq}. Here, we show the ratio (solid diamonds) between $\Delta F_Q=|F_Q(g)_\mathrm{2LL}-F_Q(g)_\mathrm{LLL}|$, the difference in Fisher information obtained with the LLL approximation and that obtained with two levels at the critical frequency $\Omega_c^{(\mathrm{LLL})}$ obtained with a LLL calculation, and the Fisher information $F_Q$ obtained using a two Landau levels approximation at this same critical frequency. Also, we show the same ratio (empty diamonds) for the case involving two Landau levels and three Landau levels. These results show vast disagreement in the Fisher information calculated with the LLL approximation with respect to that one obtained with higher Landau levels, which in principle are ``closer'' to the true value of the quantum Fisher information at that rotation frequency. This disagreement even holds for values of the interaction strength $g$ which are typically regarded as pertaining to the weakly interacting regime for which the LLL approximation is considered valid.

In particular, it is worth noting that for the value $g=6/N$, i.e. $g=0.5$ for the results in Figs.\ref{fig:FracCF} and \ref{fig:FracFq}, the LLL approximation predicts a critical rotation frequency with a relative error of less than 6\% with respect to the rotation frequency calculated with two Landau levels, however, the LLL prediction of the quantum Fisher information has a relative error of more than 8000\%. This is directly related to the fact that for these values of the interaction strength, the quantum Fisher information obtained with two Landau levels is a very sharp resonance as a function of $g$, whereas that obtained with LLL approximation remains approximately broad across these values of the interaction strength.


Notably, as the interaction strength is decreased much below the value of $g=6/N$, we recover the lowest Landau level regime again. In particular we observe that for an interaction strength of about $g\sim 0.2$ in the two Landau level approximation, we get very similar results to  those originally considered using the LLL approximation in \cite{dagnino}, i.e. a ``bat-like'' state with a broad quantum Fisher information curve. These results are shown in Fig.\ref{fig:sweetspot}. This ``bat-like'' state is extremely interesting from the point of view of quantum metrology due to three main advantageous characteristics among other ones. First, the phase precision $\Delta \phi$ scales as $\sqrt{2}/\sqrt{N(N+2)}$ and thus, for large number of particles it is comparable to the precision scaling of a ``cat'' state $\sqrt{2}/N$ which is known as Heisenberg-limited precision. This sub-shot-noise limited precision for the ``bat-like'' state is already present even for modest number of particles $N\sim 10$. Second, it has been shown \cite{Cooper2010a} that in the more realistic scenario where we have particle losses, the ``bat'' state retains a much higher sensitivity than the ``cat'' state, making it a robust choice in interferometric schemes. Finally, from the experimental point of view, the width of the Fisher information curve for the "bat-like" state is broader in the rotation rate $\Omega$. This is very desirable since it allows for a greater margin of error when trying to nucleate this entangled ground state.

These findings are exciting for metrology, not only because we can still use this system to create a robust ``bat-like'' state but we could also tune from ``N00N-like'' states to ``bat-like'' ones by only changing the interaction strength, which could be achieved in experiments using Feshbach resonances \cite{Cornish}. Tuning the system in order to get a ``N00N-like'' state can be challenging due to the fact that the Fisher information is a very sharp resonance for this case and our predictions for the critical frequency $\Omega_c$ using two levels cannot exactly pinpoint the true critical frequency. However, we can always estimate a small frequency window where we expect to find the true critical frequency by calculating $\Omega_c$ using three and four levels and extrapolating our results to predict the many-level result. The quantum Fisher information at the calculated critical frequency for each approximation is shown in Fig. \ref{fig:Fq}. These results show that the two level approximation is expected to give very accurate predictions for the Fisher information of the full many-body ground state at the \emph{true} critical frequency. 

According to our simulations for small number of particles (N=4\ldots 12) using two Landau levels, the width of the Fisher information for a fixed number of particles as calculated using the left width of the curve at half the maximum decreases with increasing interaction strength in an exponential manner, and follows approximately the same curve for each number of particles considered in our simulations, going typically from broad value of $5\times 10^{-3}$ for $g=(6/N)\times 0.4$ (``Bat-like'') to value of the order of $5 \times 10^{-5}$ for $g=(6/N)$ (``Cat-like''). Similarly, for a fixed value of $g$, the width also decreases exponentially as the number of particles is increased but the gradient is dependent on the exact value for $g$.

The quantum Fisher information at the calculated critical frequency for each approximation is shown in Fig. \ref{fig:Fq}. These results show that the two level approximation is expected to give very accurate predictions for the Fisher information of the full many-body ground state at the \emph{true} critical frequency. As Fig. \ref{fig:Fq} already suggests, we have found that the form of the entangled state and the shape of the quantum Fisher information do not change substantially when going from the two-level approximation to a three-level one for all values of the interaction strength considered in our calculations. In other words, when the shape of the entagled state given by a two-level calculation is ``bat''-like (broad Fisher information curve), it remains ``bat''-like using three levels and the converse is also true for a ``cat''-like shape (sharp Fisher information curve).   



\section{Conclusions} 

Rotating matter waves in anisoptropic potentials could prove to be an exciting system for generating useful entangled states for metrology schemes and studying quantum phase transitions. However, caution needs to be exercised when choosing experimental parameters. While theoretical studies show that the LLL approximation predicts the critical frequency well, a more in-depth analysis shows that it predicts other features very poorly. Indeed the bat-like state predicted at the critical frequency by LLL is seen to be more N00N-like and has a sharp resonance in its Fisher information meaning that it is unlikely to be practical for metrology. However we have been able to identify another parameter regime that generates an entangled state with the features we would want. This system has a rich structure and may prove to be a promising tool for engineering a range of useful entangled states. The next step is to consider the dynamics of the system and a detailed scheme for demonstrating a proof-of-principle quantum-limited gyroscope.

\begin{acknowledgments}
This work was supported by Consejo Nacional de Ciencia y Tecnolog\'ia (CONACyT, Mexico) and Secretar\'ia de Educaci\'on P\'ublica (SEP, Mexico), and the Defence Science and Technology Laboratory (UK).

\end{acknowledgments}



\end{document}